\documentclass[11pt,a4paper]{article}
\pdfoutput=1
\usepackage{jheppub}
\def\Bbb{\mathbb}
\usepackage{amsfonts}
\def\BZ{\mbox{$\Bbb Z$}} 
\def\BC{\mbox{$\Bbb C$}} \def\BP{\mbox{$\Bbb P$}}

\catcode`@=11 \@addtoreset{equation}{section} \catcode`@=12

\newcommand{\sid}{\begin{eqnarray*}}
\newcommand{\sidd}{\end{eqnarray*}}

\title{Inverse algorithm and M2-brane theories}
\author[a]{Siddharth Dwivedi}
\author[a,b]{and P. Ramadevi}
\affiliation[a]{Department of Physics,Indian Institute of Technology Bombay,\\
Mumbai 400 076, India}
\affiliation[b]{Center for Quantum Spacetime,\\
Sogang University, Seoul, S.Korea}
\emailAdd{siddharth@phy.iitb.ac.in}
\emailAdd{ramadevi@phy.iitb.ac.in}
\abstract{Recent paper arXiv:1103.0553  studied the quiver gauge theories on coincident
$M2$ branes  on a singular toric Calabi-Yau 4-folds which are complex cone over
toric Fano 3-folds. There are  18 toric Fano manifolds but only 14 
toric Fano  were obtained from the forward algorithm.
We attempt to systematize the inverse algorithm which helps in 
obtaining quiver gauge theories on $M2$-branes
from the toric data of the Calabi-Yau 4-folds. In particular, we obtain 
quiver gauge theories on coincident 
$M2$-branes corresponding to the remaining 4 toric Fano 3-folds.
We observe that these quiver gauge theories cannot be given a dimer 
tiling presentation.  }
\keywords{AdS-CFT Correspondence, M-Theory}
\arxivnumber{1108.2387}
\begin{document}
\maketitle
\flushbottom

\section{Introduction}
Starting  with the works of Bagger-Lambert \cite{bag}, Gustavsson \cite{gus}, Raamsdonk \cite{ram} and Aharony-Bergman-Jafferis-Maldacena (ABJM)
 \cite{abjm}, we see interesting developments 
in the last three years  between supersymmetric Chern-Simons  
gauge theory on  coincident $M2$ branes at the tip of Calabi-Yau 4-folds  
and their string duals. For a nice review, see ref. \cite{kleb}.

For a class of the supersymmetric Chern-Simons theories which can 
be represented by a quiver diagram, $AdS_4/CFT_3$ correspondence 
has been studied \cite{mar}. The Calabi-Yau 4-fold toric data can 
be obtained for the quiver Chern-Simons theories
by a procedure called \emph{forward algorithm}. This approach 
was initially studied  to obtain toric data of Calabi-Yau 3-folds 
from 3+1-dimensional quiver supersymmetric theories
\cite{han1}. Further, if the quiver data can admit dimer 
tilings \cite{ken,he1}, then the toric data can be obtained from 
the determinant of Kastelyne matrix.
Generalising the forward algorithm/dimer tiling   procedure 
to 2+1-dimensional quiver Chern-Simons theories resulted in obtaining toric data of 
many Calabi-Yau 4-folds \cite{han3,yama,han4,han5,han6,han7a,han7b,han8}.

In the recent paper \cite{han8}, toric data, genus, 
second Betti number of 18 toric Fano 3-folds have been  tabulated. We 
believe that there must be at least one quiver Chern-Simons theory 
corresponding to every toric Calabi-Yau 4-fold which are
complex cones over these Fano 3-folds. Using the forward algorithm 
and tilings \cite{han8},
the toric data of 14 Fano 3-folds were obtained from the corresponding 
quiver Chern-Simons theories. Finding a quiver Chern-Simons 
corresponding to the 
remaining four toric Fanos ${\BP}^3,{\cal B}_1,{\cal B}_2,{\cal B}_3$
is a challenging problem which we try to attempt by systematizing
the \emph{inverse algorithm}.
 
Reversing the procedure of the
forward algorithm,  called \emph{inverse algorithm}, 
should result in obtaining quiver gauge theories from the
toric data. As already pointed out in the context of 
Calabi-Yau 3-folds \cite{han1},
the inverse algorithm has ambiguities which we will detail 
in section 2. Considering the toric Calabi-Yau 3-folds 
as embeddings inside the non-cyclic orbifolds 
${\BC}^3/({\BZ}_n \times {\BZ}_m)$ and performing
partial resolutions, the matter content and the superpotential $W$ of 
the 3+1 quiver theories were obtained \cite{han1}. However, the
adjoint matter fields could not be explained by this method.
The adjoint fields appear naturally in the algebraic approach
\cite{tapo} which involves 
matching matrix corresponding to dimer tiling.
 
Exhaustive works \cite{han3}-\cite{han8} show that all the studied 
quiver Chern-Simons theories corresponding to the Calabi-Yau 4-folds
admit dimer tiling. Further, using higgsing \cite{han7b} of 
matter fields on a known $G$-node quiver which admits tiling, 
($G-1$) node quiver and their corresponding toric data were 
obtained. In fact, the approach \cite{tapo} can be extended 
to $2+1$-dimensional
quiver gauge theories giving the results in ref. \cite{han7b}.
All the
quiver theories before or after higgsing can be
represented as tiling. Unfortunately  ${\BP}^3,{\cal B}_1,
{\cal B}_2,{\cal B}_3$ toric data have not been obtained 
from the higgsing approach.
So, we believe that these four Fano 3-folds
may not give quivers admiting tiling description. It is also 
not clear whether we can perform partial resolution of
an abelian orbifold of ${\BC}^4$ \cite{han1} 
and obtain toric data of these Fano 3-folds. 

One of the crucial step in the inverse algorithm is to 
fix the F-term and D-term charge assignments corresponding 
to the toric data. In this work, we try to understand the 
pattern of the  F-term and the D-term charges for the 14 Fano 3-folds
whose quivers are known. With this pattern identification, 
we propose (see ansatz in section 4.1) a form for these charges. 
Then, the rest of the sequence of the inverse algorithm can be performed to give the quiver data.

The plan of the paper is as follows: In section 2, we 
briefly review the forward and the inverse algorithm.
In section 3, we first
review the inverse algorithm of ${\BC}^4/{\BZ}_2$
which closely resembles Fano ${\BP}^3$ and then derive
the quiver and the  mesonic moduli space Hilbert series for Fano
${\BP}^3$. In section 4, we first
review inverse algorithm for 
Fano ${\cal B}_4$ whose quiver is known from forward algorithm.
This helps in understanding the charge assignments 
for the other three Fanos ${\cal B}_1, {\cal B}_2, {\cal B}_3$.
We then present the details of quiver and Hilbert series 
for these three Fanos in later subsections.
Finally, we summarize in section 5.
\section{Toric data ${\cal G} \rightarrow$ quiver Chern-Simons}
We will briefly discuss the inverse algorithm which is used to obtain
quiver gauge theories from the toric data ${\cal G}_{4\times c}$ describing the Calabi-Yau
(CY) 4-folds. Here $c$ denotes the number of points (including multiplity of points)
in the toric diagram.  Unlike the forward algorthim, which starts
from the quiver data and superpotential $W$  giving a unique toric data, the inverse 
algorithm is non-unique. That is, there can be many quiver gauge theories
possible from the inverse algorithm. Besides this non-uniqueness, there 
are many subtle ambiguities in choosing the toric data. They are:
\begin{enumerate}
\item Two toric data ${\cal G}$
 and ${\cal G}'$ are equivalent if they are related by any
$GL(4,{\BZ})$ transformation ${\cal T}$, that is, ${\cal G}={\cal T}.{\cal  G}'$. This 
leads to a huge pool of possible nullspace of ${\cal G}$ - namely, the charge matrix 
$Q_{(c-4)\times c}$ satisfying $Q .{\cal G}^t=0$ can be many.
\item The multiplicity of the toric points gives the toric data
with repeated columns but they represent the same Calabi-Yau 4-folds.
Usually, it is not clear which points with what
multiplicity in the toric diagram to be taken. 
We could start with no multiplicity of all the
toric points and if we end with exotic or insensible quivers, we 
could try putting multiplicity of some toric points. This is definitely
very tedious.
\end{enumerate}
We try to resolve some of these ambiquities by understanding the
pattern of the matrix $Q_{(c-4)\times c}$ of the 14 toric Fano 3-folds derived
from  the forward algorithm \cite{han8}. We can obtain the steps of the 
inverse algorithm by reversing the sequence of steps in the forward algorthm. 
\subsection{Forward Algorithm}
We will now recapitulate the essential aspects of the forward algorithm
where one starts with a ${\cal N}=2$ Chern-Simons (CS) quiver gauge theory
and superpotential $W$. For toric quivers, there are  
$N_T$ terms in $W$ with each matter field appearing only in two terms 
with opposite signs.
\begin{enumerate}
\item From the quiver data represented as a quiver diagram, we
know the number of 
gauge groups $G$ (number of nodes), CS levels
$k_a$ for each node and
$m$ number of bi-fundamental matter fields and 
adjoints fields $X_i$'s. 
From this diagram,  we can write the quiver charge matrix elements $d_{ai}$ 
where the index $a=1,2,\ldots G$ and $i=1,2, \ldots m$. 
In the CS quivers, besides $\sum_{a}d_{ai}=0$, we also 
require that the CS levels $k_a$'s have $GCD(\{k_a\})=1$
and
$\sum_{a=1}^G k_a=0$ (Calabi-Yau requirement). Using the above
conditions and the
equation (moment map) of $U(1)^G$ abelian CS quiver gauge theories,
\begin{equation}
 \mu_a(X)=\sum_i d_{ai}|X_i|^2=k_a \sigma~, \label{mome}
\end{equation}
where $\sigma$ is the scalar component of the vector superfield $V_a$,
we can obtain only ($G-2$) $D$-term equations giving a projected 
charge matrix $\Delta_{(G-2) \times m}$ from the matrix elements $d_{ai}$.
That is, the matrix elements of projected charge matrix satisfies
\begin{equation}
 \sum_i \Delta_{bi}|X_i|^2=0~,\label{proj}
\end{equation}
where $b=1,2,\ldots (G-2)$.
For example, take a $G=3$ node quiver with the 
the CS levels $k_1,k_2,-(k_1+k_2)$
where $GCD(k_1,k_2)=1$. Further the eqns.(\ref {mome}, \ref{proj}) 
suggests that the projected charge is a single row matrix 
whose elements are given by
\begin{equation}
\Delta_i=k_2 d_{1i}-k_1 d_{2i}~. \label{pchar}
\end{equation}
\item From the F-term constraint equation
$\partial W/\partial X_i=0$, we can obtain relations between matter fields.
Introducing ($G+2$) fields $v_r$'s, we can incorporate the $F$-term constraints
in the matrix $K_{ir}$ which relates the matter fields to $v_i$ in the
following way:
\begin{equation}
X_{i}=\prod_r v_r^{K_{ir}}
\end{equation}
The dual of the $K$-matrix satisfying $K.T \geq 0$ (all entries of the matrix $K.T$ are non-negative) will give a
matrix $T_{(G+2) \times c}$ where $c$ gives 
the number of GLSM sigma model fields $p_{\alpha}$'s. 
\item From the $K$ and $T$, we can write
a matrix $P=K.T$. The entries of all these matrices $K$, $T$, $P$ are integers.
For quiver theories which can admit tiling, one can read off $P$-matrix
from $W$. The $P$-matrix relates the matter fields to GLSM  $p_{\alpha}$ fields
as
\begin{equation}
X_{i}=\prod_{\alpha} p_{\alpha}^{P_{i\alpha}}~.\label{mgls}
\end{equation}
The kernel of the $T$ as well as
$P$ ($T.Q_F^t= P.Q_F^t=0$) will give 
the ${Q_F}_{(c-G-2)\times c}$ charge matrix. 
\item From the relation (\ref {mgls}), we can obtain the
baryonic charge matrix ${Q_D}_{(G-2)\times c}$ elements from the projected charge 
(\ref {proj}) matrix elements as follows:
\begin{equation}
 \Delta_{bi} =\sum_{\alpha} P_{i \alpha} (Q_D)_{b\alpha}~. \label {qdchar}
\end{equation}
\item The total $Q$
\begin{equation}
Q_{(c-4)\times c}= 
\left(\begin{matrix} 
{Q_F}_{(c-G-2)\times c}\cr
{Q_D}_{(G-2)\times c}
\end{matrix}\right)~~, 
\end{equation}
whose kernel $Q.{\cal G}^t=0$ gives the toric data 
${\cal G}_{4 \times c}$.
\end{enumerate}
\subsection{Inverse Algorithm}
Now, we can reverse the sequence and try to obtain the quiver CS theory 
on $M2$ branes at the tip of singular CY 4-folds described by ${\cal G}$. There
are additional data of toric Fano which are useful to handle some 
of the ambiguities we had enumerated.
\begin{enumerate}
\item For the toric Fano 3-folds, ${\cal G}$ can
 be written in a form where
the symmetry of the corresponding CY 4-fold $SU(4)^{i_1} \times SU(3)^{i_2}\times SU(2)^{i_3}
\times U(1)^{i_4}$ is seen as the simple roots
along the rows of ${\cal G}$. As the rank of CY is 4, we require $3i_1+2i_2+i_3+i_4=4$.
\item The second betti number $b_2$ of the toric Fano 3-fold is
related to the number of external points (E) in the toric diagram  as: $b_2=E-3$.
Further the number of baryonic symmetries for the Fano is: $b_2-1=E-4$.
This fixes that the number of rows in the $Q_D$ matrix must be 
\begin{equation}
b_2-1=G-2~.\label {node}
\end{equation}
\item With the $GL(4,{\BZ})$ freedom, it is not obvious as to what ${\cal G}'$ 
we have to choose and find the nullspace $Q$ satisfying $Q.{\cal G}'^t=0$. 
However, from the eqn.(\ref {node}), 
we know how many rows must represent $Q_D$ matrix. Equivalently, we must 
find a CS quiver with $G=b_2+1$.
\item We try to understand the pattern of $Q$ for a toric Fano  with same $b_2$ 
obtained from forward algorithm and implement the same for the missing Fano 3-folds. 
Besides $b_2$, the symmetry of the Fano also suggests how to incorporate the
charge assignments in the $Q$ matrix. Further, the multiplicity of points
in the toric diagram is also suggested by this pattern identification. 
Following this methodology, we could guess a form for $Q_F$. Using 
$Q_F$, we determine $T$ and hence $K$.
\item The number of rows of the matrix $K$ gives the number of matter fields.
From $K$, we find a relation between matter fields which are 
supposed to be $F$-term constraints. For the $G=(b_2+1)$ node  quiver with the 
F-term constraints on the matter fields, we try to  reconstruct all possible 
toric quiver superpotential $W$. Then we can draw the quiver diagram where 
the terms in $W$ must denote closed cycles in the quiver diagram.
\item Using the pattern for $Q_D$ from the forward algorithm for the 
known Fano, we can infer $Q_D$ charge assignment pattern for other Fano 3-folds.
Further, the $Q_D$ must satisfy projected quiver charge (\ref {qdchar}).
This is the non-trivial part but for small $G$,
it is not difficult to find the choice of the
CS levels which will give (\ref {pchar}) satisfying eqn. (\ref {qdchar}).
\end{enumerate} 
In the following two sections, we will first 
work out the inverse algorithm 
for two Calabi-Yau 4-folds  whose quivers are known from the tiling/forward
algorithm.  This will help to undersand the pattern in choosing the
charge matrix $Q_F$ and $Q_D$ for the unknown Fanos ${\BP}^3,{\cal B}_1,{\cal B}_2,
{\cal B}_3$. Using this pattern, we then perform the inverse algorithm for the 
missing Fano 3-folds and obtain the corresponding toric quiver CS theories.
We also compute the Hilbert series and the R-charge assignments.
\section{M2-brane theories from CY 4-folds with $b_2=1$}
From the dimer tilings, we know that the number of gauge-groups is $G=2$
for the  quiver gauge theory corresponding to orbifolds of ${\BC}^4$
(${\BC^4}/{\BZ}_k$) which from eqn. (\ref {node}) implies $b_2=1$.
The toric Fano 3-fold ${\BP}^3$ also has $b_2=1$.
Both  ${\BC^4}/{\BZ}_k$ and ${\BP}^3$ have same number of external points in 
the toric diagram and hence zero baryonic symmetry: $b_2-1=0$. That is,
$Q_D$ is zero. So, we expect to obtain a $G=2$ node quiver for the toric Fano 
${\BP}^3$ using inverse algorithm. For these Calabi-Yau 4-folds whose
$Q_D=0$, we can directly take the matrix $T={\cal G}$ because
$Q_F.T^t= Q_F.{\cal G}^t=0$. Adding multiplicity of points in the
toric diagram will result in repetition of some columns of the $T$-matrix 
which will not alter the $K$-matrix. So, 
 adding multiplicity of points in
the toric diagram will not change the quiver data for $b_2=1$ CY
4-folds. We will first implement inverse algorithm 
for the orbifold  ${\BC}^4/{\BZ}_2$ as a warm-up exercise and then 
obtain $2$-node quiver for ${\BP}^3$ Fano 3-fold.
\subsection{${\BC}^4/{\BZ}_2$}
Let us the take the toric data ${\cal G}$ for ${\BC}^4/{\BZ}_2$
obtained from the tiling approach (see eqn.(3.2) in ref.\cite{han7a}):
{\small
\begin{equation}
{\cal G}=\begin{pmatrix}1 & 1 & 1 & 1\cr -1 & 0 & -1 & 0\cr 0 & -1 & -1 & 0\cr 0 & 0 & 2 & 0
\end{pmatrix}
\end{equation}
}
As $Q_D=0$, we can take $T={\cal G}$. The matrix $K$ from $T$ will be
\begin{equation}
\begin{pmatrix}0 & 0 & 0 & 1\cr 0 & 0 & -2 & -1\cr 0 & -2 & 0 & -1\cr 2 & 2 & 2 & 1
\end{pmatrix}
\end{equation}
The row index of the $4 \times  4$ square matrix $K$ indicates that the 
number of matter fields in the quiver theory is $4$ confirming the known
data for ${\BC}^4$ orbifolds. Also, being a square matrix, it is not possible
to find $F$-term constraint equations. From the matter field content,
one can try to construct all possible connected quiver diagrams
and the loops in the quiver diagram will give gauge invariant 
terms in superpotential $W$. The information about the Chern-Simons level $k_a$
is inferred from the matrix $P=K.T$ which for this case is given as:
\begin{equation}
P = \begin{pmatrix}0 & 0 & 2 & 0\cr 0 & 2 & 0 & 0\cr 2 & 0 & 0 & 0\cr 0 & 0 & 0 & 2
\end{pmatrix}
\end{equation}
For ${\BC}^4$, the $P$ matrix from inverse algorithm
will turn out to have entries zero or 1. So, the 
entries for ${\BC}^4/{\BZ}_2$ which are $2$ or $0$
indicates that one of the nodes of the 2-node quiver diagram must have 
level $k_1=2$ and the other node by Calabi-Yau requirement has level $k_2=-k_1=-2$. 
Inferring the CS levels from $P$-matrix is only applicable for
CY 4-folds whose $Q_D=0$. The $Q_F$ satisfying $P.Q_F^t=0$ is also trivial. 

Hence from the inverse algorithm for ${\BC}^4/{\BZ}_2$, there can be
three  possible 2-node quivers with four matter fields and levels $k_1=-k_2=2$. 
They are:
\begin{enumerate}
\item ABJM theory with 4 bi-fundamental matter fields $X_{12}^i$ and $X_{21}^i$
where $i=1,2$ and $W=Tr[\epsilon_{ij}X_{12}^1X_{21}^iX_{12}^2X_{21}^j]$
whose abelian $W=0$. 
\item 2-node quiver with two adjoints $\phi_2^1, \phi_2^2$
at the same node (say node 2) and two bifundamentals $X_{12},X_{21}$. Here 
$W=Tr[X_{12}[\phi_2^1,\phi_2^2]X_{21}]$ whose abelian $W$ is again zero.
\item 2-node quiver with two bifundamentals $X_{12}$,$X_{21}$
and adjoints $\phi_1$ and $\phi_2$ at two different nodes with trivial $W=0$.
\end{enumerate} 
It is important to realise that {\it
it is not possible to do forward algorithm
for quivers whose abelian superpotential $W$ is zero}. Particularly,
we cannot obtain the matrix $K$ for these superpotentials. Fortunately, 
the first two quivers admit dimer tiling presentation \cite{han3}. 
So, we can
obtain toric data ${\cal G}$ directly from the 
determinant of the Kastelyne matrix. We would also like to point out that the
$P$ matrix obtained from $W$ is same for ${\BC}^4$ and 
${\BC}^4/{\BZ}_2$ indicating that the $P$ matrix does not give information about 
the Chern-Simon levels. From inverse algorithm, we actually find the 
$P$-matrix entries  change with change in CS levels.

Comparing the tiling approach and the inverse algorithm, we infer that
third quiver is not allowed for ${\BC}^4/{\BZ}_2$.
We will see a similar situation arising in the
following section on Fano ${\BP}^3$. 
\subsection{Fano ${\BP}^3$ theory}
The symmetry group of the Calabi-Yau 4-folds constructed as 
complex cone over Fano ${\BP}^3$ is $SU(4) \times U(1)$.
Further $b_2=1$ implies $G=2$-node quiver and $Q_D=0$. Therefore,
the matrix $T={\cal G}$ where the toric data ${\cal G}_{4 \times c}$
respecting the symmetry is:
\begin{equation}
\begin{pmatrix}1 & 1 & 1 & 1 & 1\cr 1 & -1 & 0 & 0 & 0\cr 0 & 1 & -1 & 0 & 0
\cr 0 & 0 & 1 & -1 & 0\end{pmatrix}
\end{equation}
Here the number of points in the toric diagram
$c=5$ and hence $(Q_F)_{(c-4) \times c}$ will be a single row 
with five entries. The first four columns in ${\cal G}$ are the 
external points and
the last column denotes the internal point in the toric diagram.

For the given symmetry $SU(4) \times U(1)$, we can always choose
the fermionic charge assignment $Q_F=(aaaab)$ where $a,b$ are
integers. Further $Q_F.{\cal G}^t=0$ implies $4a+b=0$.
 Conversely, the charge $Q_F$
with four entries same reflects that the Calabi-Yau 4-fold 
has $SU(4)$ symmetry. The possible choice of $a,b$ in $Q_F$ is
\begin{equation}
Q_F=(1,1,1,1,-4)~. \label{qfchar}
\end{equation}
The matrix $K$ such that $(K.T) \geq 0$ turns out be again 
a $4 \times 4$ matrix.
 \[\begin{pmatrix}1 & -1 & -2 & -3\cr 1 & -1 & -2 & 1\cr 1 & -1 & 2 & 1\cr 1 & 3 & 
2 & 1\end{pmatrix}\]
So, for the toric data for ${\BP}^3$, the
 $K$-matrix
implies that the number of matter fields must be again 4 
which is consistent with eqn. ({\ref {node}).

As the toric data ${\cal G}$ for ${\BP}^3$ is not related to 
toric data of the orbifold
${\BC}^4/{\BZ}_k$ by $GL(4,{\BZ})$, we would expect that the 
corresponding $K$-matrices are not related by the
$GL(4,{\BZ})$ and it is indeed true. Therefore, the quiver 
for ${\BP}^3$ toric data has to be different
from the quiver for ${\BC}^4/{\BZ}_k$
toric data.

The matching matrix $P=K.T$ is given by:
\begin{equation}
P= \begin{pmatrix}0 & 0 & 0 & 4 & 1\cr 0 & 0 & 4 & 0 & 1\cr 0 
& 4 & 0 & 0 & 1\cr 4 & 0 & 0 & 0 & 1\end{pmatrix}
\end{equation}
We see that the non-zero entries in the first four columns
corresponding to the external points of ${\BP}^3$ toric diagram is 4.
Following our results on ${\BC}^4/{\BZ}_2$,
it appears that the  levels of the quiver CS theory with 
four matter fields will be $k_1=-k_2=4$. 

The quiver corresponding to ${\BP}^3$ must  be a 
2-node quiver with 4 matter fields 
with levels $k_1=-k_2=4$, which can again have three possibilities.
Two of the quivers corresponds to 
${\BC}^4/{\BZ}_4$ from the tiling approach \cite{han3}.
 By the elimination process, 
we claim that the quiver drawn in figure~\ref{fig:afano} represents the
quiver gauge theory on the coincident $M2$ branes at the tip of the
 singular Calabi-Yau, which is complex cone over Fano ${\BP}^3$
 with superpotential $W=0$. 
\begin{figure}
\centerline{\includegraphics[width=4in]{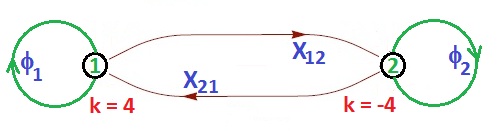}}
\caption{Quiver Diagram for Fano $P^3$}
\label {fig:afano}
\end{figure}
All the studied quivers with $N_T$ terms in $W$, $G$ nodes
and $E$ edges satisfied $N_T-E+G=0$ and hence could be drawn as 
tiling on a two-torus \cite{han3, han4, han5,han6,han7a,han7b}.
This quiver satisfies $N_T-E+G=-2$. It will be interesting to see whether this
quiver could admit a $3d$ tiling which will help to obtain 
the Kastelyne matrix and the toric data \cite{sang,sang1}.
We also believe that this quiver must be obtainable by the 
Higgsing of some quiver gauge theory, which does not
admit tiling presentation, with three gauge group nodes.

Taking the $Q_F$ (\ref {qfchar}) for the toric data ${\cal G}$, we will
work out the Hilbert series of the mesonic moduli space. As 
this Fano has only one $U(1)$ charge, which can be taken as R-charge,
the Hilbert series must have the following expected form:
\begin{equation}
g^{\rm mes}(t;X)=\frac{1+(g-2)t+(g-2)t^2+t^3}{(1-t)^4}~. \label {hilbert}
\end{equation}
where $X$ denotes Fano 3-fold of genus $g$.
Following Ref. \cite{han8}, we will take
the $R$-charge fugacity of the four external points 
as $s_1$ and for the internal point as 1. 
Then the Hilbert series (\ref{hilbert}) for this case will come out to be:
\begin{eqnarray}
 g^{mes}(s_1,{\BP}^3)&=& 
\oint_{|z|=1} \frac{dz}{2\pi i z} \frac{1}{(1-s_1z)^4 (1-z^{-4})}\nonumber\\
~&=& \frac{(1+31z^4+31z^8+z^{12})}{(z^4-1)^4}\vert
_{z=1/s_1}
\end{eqnarray}
Excluding the poles on the boundary of the contour, 
we indeed get the expected form (\ref {hilbert}) with 
the correct genus $g=33$ confirming that the 
charge assignment and the toric data we considered 
(with no multiplicity) is correct. 

We will now try to understand the inverse algorithm for 
other Fano 3-folds whose $b_2=2$ in the following section.
\section{M2-brane theories from CY 4-folds with $b_2=2$}
There are four toric Fano 3-folds ${\cal B}_i$ whose second Betti number $b_2=2$.
They have $E=5$ external points and one internal point in the toric diagram. 
From the tiling/forward algorithm, Fano ${\cal B}_4$ toric data 
and hilbert series were derived from a $G=3$ node quiver gauge theory.
As ($b_2-1)=(E-4)=(G-2)=1$ for all these Fano ${\cal B}_i$'s,
we expect to obtain $G=3$ node quivers from the inverse algorithm.
Further, the number of baryonic symmetries is $(b_2-1)=1$ 
for these Fano 3-folds. So, $Q_D$ matrix will be a single row matrix.

In order to understand the pattern of $Q_F$ and $Q_D$ matrix,
we will first do the inverse algorithm for the Fano ${\cal B}_4$
and obtain the same $3$-node quiver known from the tiling/forward algorithm.
Then, we will repeat the similar $Q$-charge pattern
for the other three Fano 3-folds and obtain their corresponding
3-node quiver gauge theories.

\subsection{Fano $\mathcal B_4$ theory}
The CY 4-fold obtained from the complex cone over ${\cal B}_4$ has
$SU(3)\times SU(2)\times U(1)$ symmetry.
The toric data ${\cal G}$ which reflects this symmetry is
\begin{equation}
G=\begin{pmatrix}1 & 1 & 1 & 1 & 1 & 1\cr 1 & -1 & 0 & 0 & 0 & 0\cr 
0 & 1 & -1 & 0 & 0 & 0\cr 0 & 0 & 0 & 1 & -1 & 0\end{pmatrix}
~.
\end{equation}
For the Fano with the given symmetry, the $Q$ matrix must have 
first three columns are identical, fourth and fifth column 
to be identical. Observing the $Q$-charge pattern for the 14-toric Fano from the 
tiling/forward algorithm, we propose the following:\\
\underline{\bf Ansatz}: {\it $Q_F$ matrix columns must possess
 non-abelian symmetry whose rank is one higher than that of the
$Q_D$ matrix. Further, the ranks of the non-abelian subgroups in $Q_F$  must be atmost
the maximal rank  of the subgroups representing the symmetry
of the toric CY 4-folds.}\\
This proposal helps in fixing the multiplicity of the points in the
toric diagram as well. 
For ${\cal B}_4$,
$Q_F$ matrix must have $SU(3)\times SU(3)$ symmetry. 
That is, the integer entries of the  $Q_F$ matrix must be:
\begin{equation}
{Q_F}_{(c-G-2)\times c}=\begin{pmatrix}a_1&a_1&a_1&b_1&b_1&b_1&c_1&d_1\ldots \cr
                       a_2&a_2&a_2&b_2&b_2&b_2&c_2&d_2\ldots\cr
.&.&.&&&&&\ldots\cr.&.&.&&&&&\ldots\cr
a_{c-5}&a_{c-5}&a_{c-5}&b_{c-5}&b_{c-5}&b_{c-5}&c_{c-5}&d_{c-5} \ldots  \end{pmatrix}~,
\end{equation}
where the $\ldots$ denotes possible multiplicities of the points
in the toric diagram. The single row $Q_D$ 
entries  must have $SU(3)\times SU(2)\times U(1)$:
\begin{equation}
{Q_D}_{1 \times c}=\left(\begin{matrix}x&x&x&y&y&l&m\ldots\end{matrix}\right)
\end{equation}
so that the total $Q$ possesses the symmetry of the Fano ${\cal B}_4$.

For the given ${\cal G}$ with no multiplicity,
we are able to find a $Q_F$ and $Q_D$ obeying the above pattern:
\begin{equation}
Q=\begin{pmatrix}Q_F \cr Q_D\end{pmatrix}=\begin{pmatrix}1&1&1&-1&-1&-1\cr0&0&0&1&1&-2\end{pmatrix}
\label {chang}
\end{equation}
The $T$ and $K$ matrix for this choice of $Q_F$ is:
\begin{equation}
T=\begin{pmatrix}1 & 0 & 0 & 0 & 0 & 1\cr 1 & 0 & 0 & 0 & 1 & 0\cr 
1 & 0 & 0 & 1 & 0 & 0\cr -1 & 0 & 1 & 0 & 0 & 0\cr -1 & 1 & 0 & 0 & 0 & 0\end{pmatrix}
~;~
 K=\left(\begin{array}{c|ccccc}
        {}&v_1&v_2&v_3&v_4&v_5\\
\hline
X_{12}^{(1)}&0& 0 & 1 & 0 & 0\\

X_{12}^{(2)}&0& 0 & 1 & 0 & 1\\
X_{12}^{(3)}&
0 & 0 & 1 & 1 & 0\\
X_{23}^{(1)}&
0 & 1 & 0 & 0 & 0\\
X_{23}^{(2)}&0
 & 1 & 0 & 0 & 1\\
X_{23}^{(3)}&
 0 & 1 & 0 & 1 & 0\\
X_{31}^{(1)}&
1 & 0 & 0 & 0 & 0\\
X_{31}^{(2)}&
1 & 0 & 0 & 0 & 1\\
X_{31}^{(3)}&
 1 & 0 & 0 & 1 & 0\end{array}\right)
\end{equation}
From the $K$-matrix, we know that the number of matter fields is $9$.
We can also reconstruct toric superpotential $W$ using the $F$-term constraints
given by the $K$-matrix:
\begin{equation}
 W={\rm Tr} \left(\epsilon_{ijk}X_{12}^{(i)}X_{23}^{(j)}X_{31}^{(k)}\right)~
\end{equation}
and draw the $3$-node cyclic quiver where the matter fields
$X_{ij}^{(l)}$'s are  bifundamental fields from 
the node $i$ to the node $j$ which will determine the quiver charge matrix:
\begin{equation}
 d=
 \left(\begin{array}{c|c c c}&X_{12}^{(i)}&X_{23}^{(j)}&X_{31}^{(k)}\\ \hline
a=1&1&0&-1\\
a=2&-1&1&0\\
a=3&0& -1&1   
\end{array}\right) \label {d4char}
\end{equation}
In order to determine the CS levels $k_a$'s, we have to write 
the $P=K.T$ matrix:
\begin{equation}
P=
\left(\begin{array}{c|cccccc}&p_1&p_2&p_3&p_4&p_5&p_6\\ \hline
X_{12}^{(1)}
&1& 0 & 0 & 1 & 0 & 0\\
X_{12}^{(2)}
&0 & 1 & 0 & 1 & 0 & 0\\ 
X_{12}^{(3)}
&0 & 0 & 1 & 1 & 0 & 0\\ 
X_{23}^{(1)}
&
1 & 0 & 0 & 0 & 1 & 0\\
X_{23}^{(2)}
&
0 & 1 & 0 & 0 & 1 & 0\\
X_{23}^{(3)}
 &
0 & 0 & 
1 & 0 & 1 & 0\\
X_{31}^{(1)}
&1 & 0 & 0 & 0 & 0 & 1\\
X_{31}^{(2)}&0 & 1 & 0 & 0 & 0 & 1\\ 
X_{31}^{(3)}
&
0 & 0 & 1 & 0 & 0 & 1
\end{array}\right)~,
\end{equation}
where we have again indicated the matter fields which represent the 
row index and the GLSM $p_{\alpha}$ 
denoting the column index which will help to write these 
matter fields as products
of GLSM $p_{\alpha}$ fields. Using the $Q_D$ charge (\ref {chang}) and 
the $P$-matrix elements, we can obtain the projected charge 
$\Delta$ of the matter fields (\ref {qdchar}):
\begin{equation}
 \Delta=
\left(\begin{array}{ccc}X_{12}^{(1,2,3)}
&X_{23}^{(1,2,3)}&                  X_{31}^{(1,2,3)}\\ \hline
     1&1&-2
  \end{array}\right) \label {p4char}
\end{equation}
Substituting the projected charge and the $d$-matrix elements (\ref {d4char})
in eqn.(\ref{pchar}), we find that the CS levels have to be
\begin{equation}
 k_1=1~,~k_2=-2~,k_3=1~.
\end{equation}
Thus, using inverse algorithm with a possible choice of $Q$ matrix respecting
the ansatz  on the $Q_F,Q_D$ pattern, we have obtained the same
$3$-node quiver CS theory with $W$, CS levels and $\Delta$ charge
\cite {han8} confirming that the inverse algorithm is agreeing 
with the forward algorithm for the Fano ${\cal B}_4$. Now, we are in a position to 
extend this pattern approach for other ${\cal B}_i$ Fano 3-folds 
and obtain the corresponding quiver CS theories.
The Hilbert series of the mesonic moduli space worked out in 
ref. \cite {han8} involves taking a simple pole or irrational
pole in each of the two integration variables to the boundary
of the contour by scaling the variables. 
Evaluating the contour in the scaled variables and excluding
the poles at the boundary gives the 
form (\ref {hilbert}) agreeing with the genus $g=28$.
\subsection{Fano ${\cal B}_1$ theory}
The symmetry possessed by this
 ${\cal B}_1$ Fano is $SU(3)\times U(1)^2$. 
Using our ansatz,
we need to choose $Q_F$ to have $SU(3) \times SU(2)\times U(1)$ symmetry. 
This could not be achieved without taking multiplicity of the 
points in the toric diagram. Hence, we start with the 
following  toric data:
\begin{equation}
{\cal G}= \left(
\begin{array}{ccccccc}
1&1&1&1&1&1&1\\
 1 & -1 & 0 & 0 & 0 & 0 & 0 \\
 0 & 1 & -1 & 0 & 0 & 0 & 0 \\
 0 & 0 & 2 & -1 & 1 & 1 & 0
\end{array}
\right)
\end{equation}
where we have taken multiplicity of an external point 
which is shown as repeated columns 5 and 6 in ${\cal G}$.
The last column represents the internal point in the toric diagram.
The $Q_F$ matrix will have now 2 rows and can be taken, following
ansatz, as:
\begin{equation}
Q_F = \left(
\begin{array}{ccccccc}
 1 & 1 & 1 & -2 & -2 & -2 & 3 \\
 0 & 0 & 0 &  2 &  1 &  1 & -4
\end{array}
\right) \label{qb1}
\end{equation}
We can take a possible choice for the single row
 $Q_D$ imposing its entries so that the $Q$ matrix 
 respect $SU(3) \times U(1)^2$ symmetry:
\begin{equation}
Q_D = \left(
\begin{array}{ccccccc}
 0, & 0, & 0, & 1, & 0, & 1, & -2
\end{array}
\right)\label{qd:b1}
\end{equation}
With this $Q_F$, we can find $T$ and hence the $K$ matrix as:
\begin{equation}
T=\begin{pmatrix}1 & 0 & 0 & 2 & 0 & 0 &1\cr 2 & 0 & 0 & -1 & 0 & 2 & 0\cr 
2 & 0 & 0 & -1 & 2 & 0 &0\cr -1 & 0 & 1 & 0 & 0 & 0 & 0\cr -1 & 1 & 0 & 0 & 0 & 0 & 0\end{pmatrix}
~;~
 K=\left(\begin{array}{c|ccccc}
        {}&v_1&v_2&v_3&v_4&v_5\\
\hline
X_1&1 & 0 & 0 & 0 & 0 \\
X_2&
 1 & 0 & 0 & 0 & 1 \\
X_3&1 & 0 & 0 & 1 & 0 \\
X_4&1 & 0 & 2 & 0 & 0 \\
X_5&1 & 0 & 2 & 0 & 5 \\
X_6&1 & 0 & 2 & 5 & 0 \\
X_7&1& 2 & 0 & 0 & 0 \\
X_8&1 & 2 & 0 & 0 & 5 \\
X_9& 1 & 2 & 0 & 5 & 0
\end{array}
\right)
\end{equation}
The $K$-matrix indicates that there are $9$ matter fields $X_i$'s. 
From the $K$-matrix, we do 
find the following relations: $X_9 X_4=X_6 X_7,~X_5 X_9= X_6 X_8,~
X_5 X_7=X_4 X_8,$ and
we could construct a toric superpotential $W$ respecting
the above equation as 
$W={\rm Tr}\left[
X_1(X_5 X_9-X_6 X_8)-X_2(X_5 X_7-X_4 X_8)+X_3(X_6X_7-X_9X_4)\right]~.$\\
However, the F-term constraint 
$\partial W\over \partial X_i$ for $X_4,X_5,\ldots X_9$
is not respected by the $K$-matrix. Hence the only possible 
toric superpotential
$W$ will be a two-term superpotential with each term involving all the $9$ matter
fields:
\begin{equation}
 W={\rm Tr} \left[X_2X_5X_8\{X_1X_4X_9X_3X_6X_7-X_1X_6X_7X_3X_4X_9\}\right]~.\label {potb1}
\end{equation}
Clearly, abelian $W=0$ and hence we cannot perform forward algorithm for this
$3$-node quiver. Further $N_T-E+G=2-9+3=-4$ and hence cannot admit
tiling presentation. 

The $W$ suggests the 3-node quiver
must be as shown in figure~{\ref {fig:b1}. 
From the quiver diagram, we can obtain the $d$-matrix elements:
\begin{equation}
d_{ai} = \left(
\begin{array}{c|ccc}
    & X_i(i=1,2,3) & X_i(i=4,5,6) & X_i(i=7,8,9)     \\
\hline
a=1 &  -1 &  1 & 0  \\
a=2 & 0 &  -1 & 1  \\
a=3 &  1 & 0 & -1
\end{array}
\right)\label {d1char}
\end{equation}
To determine the CS levels of the three nodes, we 
need the $P=K.T$ matrix:
\begin{equation}
P=\left(
\begin{array}{ccccccc}
1&0&0&2&0&0&1\\
0&1 & 0 & 2 & 0 & 0 & 1 \\
 0 & 0 & 1 & 2 & 0 & 0 & 1 \\
 5 & 0 & 0 & 0 & 4 & 0 & 1 \\
 0 & 5 & 0 & 0 & 4 & 0 & 1 \\
 0 & 0 & 5 & 0 & 4 & 0 & 1 \\
 5 & 0 & 0 & 0 & 0 & 4 & 1 \\
 0 & 5 & 0 & 0 & 0 & 4 & 1 \\
 0 & 0 & 5 & 0 & 0 & 4 & 1
\end{array}
\right) \label {p1mat}
\end{equation}
Using the above matrix elements and the $Q_D$
(\ref{qd:b1}), we obtain the following $\Delta$ charge 
for the nine-matter fields in the $G=3$ node quiver:
\begin{equation}
\Delta=\left(\begin{array}{ccc}
              X_{1,2,3}&X_{4,5,6}&X_{7,8,9}\\ \hline
0& -2&2
\end{array}\right)\label {p1char}
\end{equation}
\begin{figure}
\centerline{\includegraphics[width=2in]{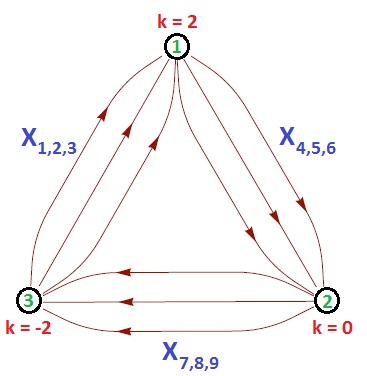}}
\caption{${\cal B}_1$ Quiver Diagram}
\label {fig:b1}
\end{figure}
Substituting $\Delta$ charge
(\ref {p1char}) and $d$ charge (\ref {d1char}) in eqn.(\ref {pchar}),
we find that the CS levels of the three nodes must be
\begin{equation}
 k_1=2~, k_2=0~,k_3=-2~.
\end{equation}
As a further consistent check with our charge assignment 
(\ref{qb1}, \ref{qd:b1}),
we will work out the Hilbert series for the mesonic moduli space 
in the following subsection.
\subsubsection{Hilbert series  for ${\cal B}_1$ Theory}
The total charge matrix for the ${\cal B}_1$ theory (\ref{qb1}, \ref{qd:b1}) is
\begin{equation}
Q = \left(
\begin{array}{c}
 Q_F \\
 Q_D
\end{array}
\right) = \left(
\begin{array}{ccccccc}
p_1&p_2&p_3&p_4&p_5&\tilde p_5&p_6\\
\hline
 1 & 1 & 1 & -2 & -2 & -2 & 3 \\
 0 & 0 & 0 & 2 & 1 & 1 & -4 \\
 0 & 0 & 0 & 1 & 0 & 1 & -2
\end{array}
\right)\label {chb1}
\end{equation}
Here, we have $c=7$ $p_{\alpha}$ fields. Let us denote the
R-charge fugacity associated with the first three 
$p_{\alpha}$'s ($p_1,p_2,p_3$) 
respecting $SU(3)$ symmetry as $s_1$
and fugacities associated with
$p_4$, $p_5$ and ${\tilde p}_5$ as
$s_2$, $s_3$ and $s_4$ respectively. 
Since the R-charge of the internal point $p_6$ is 0, the
corresponding fugacity is set to unity. Using the $Q$ matrix, 
the Hilbert series of the mesonic moduli space can be written as:
{\small
\begin{eqnarray}
g^{mes}(s_1,s_2,s_3,s_4;{\cal B}_1)&=& \oint_{|z_1|=1} 
\left(\frac{\text{$dz_1$}}{2 \pi  \text{$\mathit{i}$$z_1$} }\right)  \oint_{|z_2|=1} 
\left(\frac{\text{$dz_2$}}{2 \pi  \text{$\mathit{i}$$z_2$} }\right) 
\oint_{|b|=1} \left(\frac{db}{2 \pi  \text{$\mathit{i}$$b$} }\right)\nonumber\\
                                 ~&~&
\left\{\frac{1}{\left(1-s_1z_1\right)^3\left(1-s_2\frac{bz_2^2}{z_1^2}\right)
\left(1-s_3\frac{z_2}{z_1^2}\right) \left(1- s_4\frac{z_2b}{z_1^2}\right)
\left(1-\frac{z_1^3}{b^2z_2^4}\right)}\right\}\nonumber
\end{eqnarray}
}
Similar to the ${\cal B}_4$ integration, 
we have scaled each integration
variable so that the simple pole (preferably irrational pole) is on
the boundary. Evaluating the integration with the above scaling,
we obtain
\begin{eqnarray}
g^{mes}(s_1,s_2,s_3,s_4;{\cal B}_1)&=&
\frac{s_3s_4}{\left(s_1^5s_3^2s_4^2-1\right)^3\left(s_2^5-s_3s_4\right)} 
\left\{1+3s_1s_2^2+6s_1^2s_2^4\right.\nonumber\\
~&~&~~~\left.
+10s_1^3s_2s_3s_4 
+15s_1^4s_2^3s_3s_4+18s_1^5s_3^2s_4^2+19s_1^6s_2^2s_3^2s_4^2\right.\nonumber\\
~&~&~~\left.
+18s_1^7s_2^4s_3^2
s_4^2
+15s_1^8s_2s_3^3s_4^3
+10s_1^9s_2^3s_3^3s_4^3+6s_1^{10}s_3^4s_4^4 \right. \nonumber\\
~&~&\quad\left.+3s_1^{11}s_2^2s_3^4s_4^4+
s_1^{12}s_2^4s_3^4s_4^4\right\}
\end{eqnarray}
Let us define two new fugacities, $t_1 \equiv s_1^{1/2}s_3^{1/5}s_4^{1/5}$ and 
$t_2 \equiv s_2s_3^{-1/5}s_4^{-1/5}$. So, the Hilbert series 
for mesonic moduli space will be given as:
\begin{eqnarray}
g^{mes}(t_1,t_2;{\cal B}_1)&=&\frac{1}{\left(1-t_2^5\right)\left(1-t_1^{10}\right)^3} 
                                   \left\{1+10t_2t_1^6+15t_2t_1^{16}+3t_2^2t_1^2+6t_2^4t_1^4
\right.\nonumber\\
~&~&~~\left.
+15t_2^3t_1^8+18t_1^{10}
+19t_2^2t_1^{12}+18t_2^4t_1^{14}+10t_2^3t_1^{18}+6t_1^{20}\right. \nonumber\\
~&~&\left.
+3t_2^2t_1^{22}+t_2^4t_1^{24}\right\}~.
\end{eqnarray}
This is the expected polynomial form 
for the Calabi-Yau 4-folds with $U(1)^2$ symmetry.
This indirectly confirms that our charge assignment
(\ref{chb1}) is correct.
}
We can determine the $R$-charge assignment of the 
$p_{\alpha}$ fields following the steps in ref. \cite{han8}.
Suppose, $R_1$ and $R_2$ be the R-charges corresponding to the 
fugacities $t_1$ and $t_2$. 
That is, $t_1 = e^{-\mu R_1}$ and $t_2 = e^{-\mu R_2}$, where $\mu$ is the 
chemical potential for the R-charge. So, the volume of ${\cal B}_1$ will be given as:
\begin{equation}
V({\cal B}_1)=
\lim_{\mu\to 0}\mu^4 g^{mes}(e^{-\mu R_1},e^{-\mu R_2};{\cal B}_1)~. \label {vb1}
\end{equation}
However, $R_1$ and $R_2$ are not independent. Using the terms in 
the superpotential (\ref {potb1}) to have $R$-charge 2 implies that
the product of all the matter fields 
which in terms of $p_{\alpha}$ fields from $P$ matrix (\ref {p1mat})
($
\prod\limits_{i=1}^9 X_i = p_1^{11}p_2^{11}p_3^{11}p_4^6p_5^{12}p_6^{12}p_7^9
$)
must have $R$-charge 2 which implies that:
\begin{equation}
33R_1 + 3R_2 = 1
\end{equation}
Putting it in volume of ${\cal B}_1$ (\ref {vb1}) and
minimising it, we get $\hat R_1 = 0.02272$ and $
\hat R_2 = 0.08333$. 
The R-charge of the $p_{\alpha}$ field can be found by using:
\begin{equation}
R(p_{\alpha})
= \lim_{\mu\to 0} \frac{1}{\mu}\left[ \frac{g(e^{-\mu
\hat  R_i};
D_{\alpha})
}{g^{mes}(e^{-\mu \hat
R_i};
{\cal B}_1)}-1\right]
\end{equation}
where $D_{\alpha}$ is the divisor corresponding to the field 
$p_{\alpha}$ and $g(e^{-\mu \hat R_i};D_{\alpha})$ gives the 
associated Hilbert series evaluated at the $\hat R_i$ which 
minimises $V({\cal B}_1)$. For the given charge 
assignment (\ref {chb1}),
the associated Hilbert series for the divisor $D_1$ corresponding to the
field $p_1$ is
\begin{eqnarray}
g(s_1,s_2,s_3,s_4;D_1)&=& 
\oint_{|z_1|=1} \left(\frac{\text{$dz_1$}}{2 \pi  
\text{$\mathit{i}$$z_1$} }\right)  \oint_{|z_2|=1} 
\left(\frac{\text{$dz_2$}}{2 \pi  
\text{$\mathit{i}$$z_2$} }\right) \oint_{|b|=1} \left(\frac{db}{2 \pi  
\text{$\mathit{i}$$b$} }\right) \\
&&\left\{\frac{{(s_1z_1)}^{-1}}{\left(1-s_1z_1\right)^3
\left(1-s_2\frac{bz_2^2}{z_1^2}\right)\left(1-s_3\frac{z_2}{z_1^2}\right) 
\left(1- s_4\frac{z_2b}{z_1^2}\right)\left(1-\frac{z_1^3}{b^2z_2^4}\right)}\right\}\nonumber
\end{eqnarray}
The $SU(3)$ symmetry of the ${\cal B}_1$ requires $R(p_1)=R(p_2)=R(p_3)$. 
Substituting the fugacities $s_1,s_2,s_3,s_4$ and rewriting in terms of 
$t_1,t_2$, the $R$-charge of $p_1$ field turns out to be
$R(p_1)= 2 \hat R_1$. Similarly, computation of 
associated Hilbert series for divisor $D_4$ corresponding to field
$p_4$ gives $R(p_4)=\hat R_2$.
This method enables evaluation of $R$-charges of
all the $p_{\alpha}$ fields. In fact, $R(p_5)=R(p_6)=0$.

We know that ${\cal B}_1$ has two $U(1)$ symmetries. We have
found $R$-charge corresponding to one $U(1)$ symmetry.
The $q$-charge assignment, corresponding to the other
abelian mesonic symmetry $U(1)$, must be such that the superpotential 
terms (\ref {potb1}) are uncharged. Further the $q$-charge vector 
must be linearly independent to $Q_F,Q_D$ (\ref {chb1}). We have 
tabulated all these results in Table~\ref{tb1}.
\begin{table}[h]
\begin{center}
  \begin{tabular}{ | c | c | c | c | c | c |}
    \hline
  & $SU(3)$ & $U(1)_R$ & $U(1)_B$ & $U(1)_q$ & fugacity        \\  \hline
    $p_1$ & (1,0)       & 0.04545  & 0        & 1        &  $s_1y_1q$      \\  \hline
    $p_2$ &(1-,1)   & 0.04545  & 0        & 1        &  $s_1y_2q/y_1$  \\  \hline
    $p_3$ & (0,-1)   & 0.04545  & 0        & 1        &  $s_1q/y_2$     \\  \hline
    $p_4$ & (0,0)    & 0.08333  & 1        &-1        &  $s_2b/q$       \\  \hline  
    $p_5$ & (0,0)    & 0        & 0        &-3        &  $1/q^3$        \\  \hline
    ${\tilde p}_5$ & (0,0)    & 0        & 1        & 0        &  $b$            \\  \hline
    $p_6$ & (0,0)    & 0        &-2        & 1        &  $q/b^2$        \\  \hline
  \end{tabular}
\end{center}
\caption{\textbf{Various charges of $p_{\alpha}$ fields under global symmetry of 
${\cal B}_1$ theory}. In this table, $s_i$ are the fugacities of R-charges, 
$y_1$ and $y_2$ are weights of $SU(3)$ symmetry, 
$b$ and $q$ are fugacities of $U(1)_B$ and $U(1)_q$ symmetries respectively.}
\label{tb1}
\end{table}
\subsection{Fano ${\cal B}_2$ theory}}
The symmetry of ${\cal B}_2$ is also $SU(3)\times U(1)^2$. So, we
expect, from anatz, $Q_F$ charge assignmnet to respect  
$SU(3) \times SU(2) \times U(1)$. In order to achieve this symmetry,
we take multiplicity of two external points $p_4,p_5$ 
in the toric diagram giving the following toric data:
\begin{equation}
{\cal G}= \left(
\begin{array}{ccccccccc}
&p_1&p_2&p_3&p_4& {\tilde p}_4&p_5&
{\tilde p}_5&p_6 \\
\hline
&1&1&1&1&1&1&1&1\\
 &1 & -1 & 0 & 0 & 0 & 0 & 0 & 0\\
 &0 & 1 & -1 & 0 & 0 & 0 & 0 & 0\\
 &0 & 0 & 1 & -1 & -1 & 1 & 1 & 0
\end{array}
\right)
\end{equation}
and we choose the following 
$Q_F$ charge matrix with first three columns 
identical, fourth and sixth column identical so that the 
non-abelian symmetry $SU(3) \times SU(2)$ is obeyed: 
\begin{equation}
Q_F = \left(
\begin{array}{cccccccc}
 1 & 1 & 1 & -1 & -1 & -1 & -2 & 2 \\
 0 & 0 & 0 &  1 & -1 &  1 & -1 & 0 \\
 0 & 0 & 0 &  0 &  1 &  0 &  1 &-2
\end{array}
\right)\label {fb2}
\end{equation}
A possible choice for the baryonic charge $Q_D$ matrix with breaks the 
symmetry to $SU(3) \times U(1)^2$ is:
\begin{equation}
Q_D = \left(
\begin{array}{cccccccc}
 0, & 0, & 0, & 1, & 1, & 2, & 0, & -4
\end{array}
\right)\label {db2}
\end{equation}
From $Q_F$ charge assignment (\ref {fb2}),
we can find the $T$ and hence the $K$ matrix as:
\begin{equation}
T=\begin{pmatrix}2 & 0 & 0 & 2 & 2 & 0 & 0 & 1\cr 1 & 0 & 0 & 0 & -1 & 0 & 1 & 0\cr 
0 & 0 & 0 & -1 & 0 & 1 & 0 & 0\cr -1 & 0 & 1 & 0 & 0 & 0 & 0 & 0\cr -1 & 1 & 0 & 0 & 0 & 0 & 0 & 0\end{pmatrix}
~;~
K = \left(
\begin{array}{ccccc}
 1 & 0 & 0 & 0 & 0 \\
 1 & 0 & 0 & 0 & 2 \\
 1 & 0 & 0 & 2 & 0 \\
 1 & 0 & 2 & 0 & 0 \\
 1 & 0 & 2 & 0 & 2 \\
 1 & 0 & 2 & 2 & 0 \\
 1 & 2 & 0 & 0 & 0 \\
 1 & 2 & 0 & 0 & 4 \\
 1 & 2 & 0 & 4 & 0 \\
 1 & 2 & 2 & 0 & 0 \\
 1 & 2 & 2 & 0 & 4 \\
 1 & 2 & 2 & 4 & 0
\end{array}
\right)
\end{equation}
Here, the rows denote the matter fields $X_i$, ($i=1,2,...,12$).
From the $K$ matrix elements, we can construct a 
toric superpotential $W$ as
\begin{eqnarray}
W &=&{\rm Tr}\left( X_1X_4X_8X_{12} - X_1X_4X_9X_{11} - X_2X_5X_7X_{12}\right.\nonumber\\
~&~&\left. + X_2X_5X_9X_{10} + X_3X_6X_7X_{11} - X_3X_6X_8X_{10}\right)~.\label{potb2}
\end{eqnarray}
\begin{figure}
\centerline{
\includegraphics[width=2in]{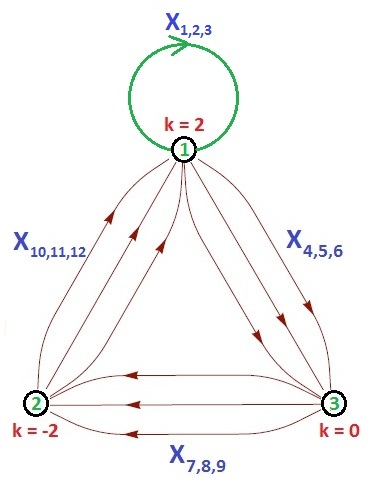}\hskip.5in 
\includegraphics[width=2in]{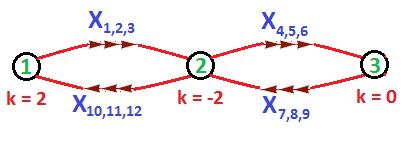}}
\caption{Two quivers (1) and (2) for the ${\cal B}_2$ toric data}
\label {fig:b2}
\end{figure}
There are $N_T=6$ terms in $W$ for the $G=3$ node quiver with $E=12$ matter fields.
Clearly, $N_T-E+G=-3$ and cannot admit tiling presentation. However,
the abelian $W$ is non-zero. So, it must be possible to do the 
forward algorithm after deducing the quiver diagram with CS levels 
from the inverse algorithm. This suggests two possible quivers 
as shown in figure~\ref{fig:b2}.

To obtain the CS levels on the three nodes of the quiver,
we need the matrix $P=K.T$:
\sid
P = \left(
\begin{array}{cccccccc}
 2 & 0 & 0 & 2 & 2 & 0 & 0 & 1 \\
 0 & 2 & 0 & 2 & 2 & 0 & 0 & 1 \\
 0 & 0 & 2 & 2 & 2 & 0 & 0 & 1 \\
 2 & 0 & 0 & 0 & 2 & 2 & 0 & 1 \\
 0 & 2 & 0 & 0 & 2 & 2 & 0 & 1 \\
 0 & 0 & 2 & 0 & 2 & 2 & 0 & 1 \\
 4 & 0 & 0 & 2 & 0 & 0 & 2 & 1 \\
 0 & 4 & 0 & 2 & 0 & 0 & 2 & 1 \\
 0 & 0 & 4 & 2 & 0 & 0 & 2 & 1 \\
 4 & 0 & 0 & 0 & 0 & 2 & 2 & 1 \\
 0 & 4 & 0 & 0 & 0 & 2 & 2 & 1 \\
 0 & 0 & 4 & 0 & 0 & 2 & 2 & 1 
\end{array}
\right)
\sidd
Using the $Q_D$ charge (\ref {db2}) and 
the $P$-matrix elements, we can obtain the projected charge 
$\Delta$ of the matter fields (\ref {qdchar}):
\begin{equation} 
\Delta=\left(\begin{array}{cccc}
              X_{1,2,3}&X_{4,5,6}&X_{7,8,9}&X_{10,11,12}\\
\hline
0& 2&-2&0
\end{array}\right)~.\label {pb2}
\end{equation}
Substituting (\ref {pb2}) in 
the charge matrix $d_1$ for the cyclic quiver in figure~\ref{fig:b2},
\begin{equation}
d_1 = \left(
\begin{array}{c|cccc}
    & X_i(i=1,2,3) & X_i(i=4,5,6) & X_i(i=7,8,9) & X_i(i=10,11,12)   \\
\hline
a=1 &  0 &  1 & 0 & -1 \\
a=2 &  0 &  0 &-1 &  1  \\
a=3 &  0 & -1 & 1 &  0
\end{array}
\right)~,
\end{equation}
or in the  charge matrix $d_2$ of the linear quiver in figure~\ref{fig:b2}
\begin{equation}
d_2 = \left(
\begin{array}{c|cccc}
    & X_i(i=1,2,3) & X_i(i=4,5,6) & X_i(i=7,8,9) & X_i(i=10,11,12)   \\
\hline
a=1 &  1 &  0 & 0 & -1 \\
a=2 & -1 &  1 &-1 &  1  \\
a=3 &  0 & -1 & 1 &  0
\end{array}
\right)~,
\end{equation}
in eqn.(\ref {pchar}), the CS levels of the 3-nodes are 
\begin{equation}
 k_1=2,  k_2=-2, k_3=0~.
\end{equation}
Incidentally, the linear and the cyclic quivers can be considered as Seiberg duals which correspond to
the same ${\cal B}_2$ toric data \footnote{We thank the referee for pointing this out.}.
To further reinforce that the charge assignments we have chosen is 
consistent, we will now do the Hilbert series of the mesonic 
moduli space for ${\cal B}_2$ toric data.\\

\subsubsection{Hilbert series evaluation for ${\cal B}_2$ Theory}
The total charge matrix for the $B_2$ theory is given by:
\begin{equation}
Q = \left(
\begin{array}{c}
 Q_F \\
 Q_D
\end{array}
\right) = \left(
\begin{array}{cccccccc}
 1 & 1 & 1 & -1 & -1 & -1 & -2 & 2 \\
 0 & 0 & 0 & 1 & -1 & 1 & -1 & 0 \\
 0 & 0 & 0 & 0 & 1 & 0 & 1 & -2 \\
 0 & 0 & 0 & 1 & 1 & 2 & 0 & -4
\end{array}
\right) \label {tcb2}
\end{equation}
The symmetry group for this theory is $SU(3) \times {U(1)}^2$. Here, we have 8 
$p_{\alpha}$ fields. Let us denote the
R-charge fugacity associated with the $p_1$, $p_2$ and $p_3$ to be $s_1$ and 
fugacities associated with
$p_4$, $\tilde p_4$, $p_5$, $\tilde p_5$ to be $s_2$, $s_3$, $s_4$ and $s_5$ respectively. 
Since the R-charge of the internal perfect matching $p_6$ is 0, the
corresponding fugacity is set to unity. Using the $Q$ matrix, the Hilbert series of 
the mesonic moduli space can be
written as:
\begin{eqnarray}
g^{mes}(s_1,s_2,s_3,s_4,s_5;{\cal B}_2)&= & 
\oint_{|z_1|=1} \left(\frac{\text{$dz_1$}}{2 \pi  \text{$\mathit{i}$$z_1$} }\right) 
 \oint_{|b|=1} \left(\frac{\text{$db$}}{2 \pi  \text{$\mathit{i}$$b$} }\right)
\oint_{|z_3|=1} \left(\frac{\text{$dz_3$}}{2 \pi  \text{$\mathit{i}$$z_3$} }\right)
\nonumber\\
~&~&
\oint_{|z_2|=1} \left(\frac{dz_2}{2 \pi  \text{$\mathit{i}$$z_2$} }\right) \left\{\frac{1}{{\left(1-s_1z_1\right)^3}
\left(1-s_2\frac{bz_2}{z_1}\right)\left(1-s_3\frac{b z_3}{z_1z_2}\right)}
\right.\nonumber\\
&~&\left.\frac{1}{
\left(1-s_4\frac{b^2z_2}{z_1}\right)\left(1-s_5\frac{ z_3}{z_1^2z_2}\right)
\left(1-\frac{z_1^2}{b^4z_3^2}\right)}\right\}
\end{eqnarray}
The Hilbert series for mesonic moduli space with the change
of variables as
$t_1=s_1\sqrt{s_4s_5}$ and $t_2=s_2s_3/\sqrt{s_4s_5}$ turns out to be
\[\frac{{t_2}^{3}\,{t_1}^{9}+3\,{t_1}^{8}+6\,t_2\,{t_1}^{7}+10\,{t_2}^{2}\,{t_1}^{6}+12\,{t_2}^{3}\,{t_1}^{5}+12\,{t_1}^{4}+10\,t_2\,{t_1}^{3}+6\,{t_2}^{2}\,{t_1}^{2}+3\,{t_2}^{3}\,t_1+1}{\left( t_2-1\right) \,\left( t_2+1\right) \,\left( {t_2}^{2}+1\right) \,{\left( {t_1}^{2}-1\right) }^{3}\,{\left( {t_1}^{2}+1\right) }^{3}}\]
which is the form expected for toric CY 4-folds with two $U(1)$ symmetries.
Suppose, $R_1$ and $R_2$ be the R-charges corresponding to the fugacities $t_1$ and $t_2$. 
That is, $t_1 = e^{-\mu R_1}$ and $t_2 = e^{-\mu R_2}$, where 
$\mu$ is the chemical potential for the R-charge. 
Similar to the volume minimisation done for ${\cal B}_1$, we  can find the value 
$\hat R_1, \hat R_2$ which minimises the volume $V({\cal B}_2)$. 
For the given $W$ (\ref {potb2}), we find that $3R_1+R_2=1/2$.
Substituting this relation in the volume of ${\cal B}_2$ and minimising it, 
we get $R_1 = R_2 = 1/8 $. Following the methods of ${\cal B}_1$, using 
the eqn. (\ref {tcb2}), the
R-charges of each $p_{\alpha}$  can be 
similarly determined which we tabulate in Table~\ref{tb2}.

\noindent 
\begin{table}[h]
\begin{center}
  \begin{tabular}{ | c  | c | c | c | c | c |}
    \hline
  & $SU(3)$& $U(1)_R$ & $U(1)_B$ & $U(1)_q$  & fugacity        \\  \hline
    $p_1$ & (1,0)       & 1/8      & 0        &   0       &  $ty_1$      \\  \hline
    $p_2$ &(-1,1)       & 1/8      & 0        &  0       &  $ty_2/y_1$  \\  \hline
    $p_3$ & (0 ,-1)     & 1/8      & 0        &   0      &  $t/y_2$     \\  \hline
    $p_4$ & (0,0)       & 0        & 1        &   0       &  $b$      \\  \hline  
    ${\tilde p}_4$ & (0,0)       & 1/8      & 1        &  0     &  $tb$     \\  \hline
    $p_5$ & (0,0)       & 0        & 2        &   1      &  $b^2q$            \\  \hline
    ${\tilde p}_5$ & (0,0)       & 0        & 0        &   -1       & $1/q$        \\  \hline
    $p_6$ & (0,0)       & 0        &-4        &   0       &  $1/b^4$        \\  \hline
  \end{tabular}
\end{center}
\caption{\textbf{Various charges of perfect matchings under global symmetry of 
${\cal B}_2$ theory}. In this table, $t$ is the fugacity of 
R-charges, $y_1$ and $y_2$ are weights of $SU(3)$ symmetry, $b$ and $q$ are 
fugacities of $U(1)_B$ and $U(1)_q$ symmetries respectively.}
\label{tb2}
\end{table}
\subsubsection{Genus  for ${\cal B}_2$ Theory}
From the fugacities of the 8 perfect matchings listed in the Table~\ref{tb2}, the Hilbert series of the mesonic moduli space can be obtained by integrating over the fugacities $z_1$, $z_2$, $z_3$ and $b$ associated with the three rows of $Q_F$ and one row of $Q_D$ respectively, which is given below:
\begin{eqnarray}
g^{mes}(t,q,y_1,y_2;{\cal B}_2)&= & 
\oint_{|z_1|=1} \left(\frac{\text{$dz_1$}}{2 \pi  \text{$\mathit{i}$$z_1$} }\right) 
 \oint_{|b|=1} \left(\frac{\text{$db$}}{2 \pi  \text{$\mathit{i}$$b$} }\right)
\oint_{|z_3|=1} \left(\frac{\text{$dz_3$}}{2 \pi  \text{$\mathit{i}$$z_3$} }\right)
\nonumber\\
~&~&
\oint_{|z_2|=1} \left(\frac{dz_2}{2 \pi  \text{$\mathit{i}$$z_2$} }\right) \left\{\frac{1}{{\left(1-ty_1z_1\right)}
\left(1-\frac{ty_2z_1}{y_1}\right)\left(1-\frac{tz_1}{y_2}\right)\left(1-\frac{bz_2}{z_1}\right)}
\right.\nonumber\\
&~&\left.\frac{1}{
\left(1-\frac{b t z_3}{z_1z_2}\right)
\left(1-\frac{b^2qz_2}{z_1}\right)\left(1-\frac{z_3}{qz_1^2z_2}\right)\left(1-\frac{z_1^2}{b^4z_3^2}\right)}\right\}
\end{eqnarray}
After doing the integration and setting the fugacities other 
than that of $U(1)_R$ charges as 1, we get the Hilbert series as:
\begin{eqnarray}
g^{mes}(t,y_1=1,y_2=1,q=1;{\cal B}_2)=\frac{1+31t^4+31t^8+t^{12}}{\left(1-t^4\right)^4}
\end{eqnarray}
This result is in the form expected for the Calabi-Yau 4-folds
(\ref{hilbert}) but it is not clear why it is not giving the genus $g=29$ \cite{han8}.
 \subsection{Fano ${\cal B}_3$ theory}
The symmetry of ${\cal B}_3$ is $SU(2)^2\times U(1)^2$. 
We take multiplicity of internal point $p_5$ as two
in the toric diagram giving the following toric data:
\begin{equation}
{\cal G}= \left(
\begin{array}{ccccccc}
p_1&p_2&p_3&p_4&p_5&
{\tilde p}_5&p_6 \\
\hline
1&1&1&1&1&1&1\\
 1 & -1 & 0 & 0 & 0 & 0 & 0\\
 0 & 0 & 1 & -1 & 0 & 0 & 0\\
 0 & 1 & 0 & -1 & -1 & -1 & 0
\end{array}
\right)
\end{equation}
Following ansatz,
we choose the $Q_F$ charge matrix with two rows 
respecting non-abelian $SU(2)^3$ symmetry: 
\begin{equation}
Q_F = \left(
\begin{array}{ccccccc}
 1 & 1 & 3 & 3 & -1 & -1 & -6 \\
 1 & 1 & 1 &  1 &  0 &  0 &  -4
\end{array}
\right)\label {fb3}
\end{equation}
A possible choice for the baryonic charge $Q_D$ matrix which breaks the 
symmetry to $SU(2)^2 \times U(1)^2$ is
\begin{equation}
Q_D = \left(
\begin{array}{ccccccc}
 0, & 0, & 0, & 0, & 1, & -1, & 0
\end{array}
\right)\label {db3}
\end{equation}
From $Q_F$ charge assignment (\ref {fb3}),
we can find the $T$ and hence the $K$ matrix as:
\begin{equation}
T=\begin{pmatrix}3 & 0 & 1 & 0 & 0 & 0 & 1\cr -1 & 0 & 1 & 0 & 0 & 2 & 0\cr 
-1 & 0 & 1 & 0 & 2 & 0 & 0\cr 0 & 0 & -1 & 1 & 0 & 0 & 0\cr -1 & 1 & 0 & 0 & 0 & 0 & 0\end{pmatrix}
~;~
K = \left(
\begin{array}{ccccc}
 1 & 0 & 0 & 0 & 0 \\
 1 & 0 & 0 & 0 & 3 \\
 1 & 0 & 0 & 1 & 0 \\
 1 & 0 & 0 & 1 & 3 \\
 1 & 0 & 3 & 0 & 0 \\
 1 & 0 & 3 & 4 & 0 \\
 1 & 3 & 0 & 0 & 0 \\
 1 & 3 & 0 & 4 & 0 
 \end{array}
\right)
\end{equation}
Here, the rows denote the matter fields $X_i$, ($i=1,2,...,8$).
From the $K$ matrix elements, we can construct a 
toric superpotential $W$ as
\begin{eqnarray}
W =(X_1X_4-X_2X_3)(X_5X_8-X_6X_7)
\label{potb3}
\end{eqnarray}
\begin{figure}
\centerline{\includegraphics[width=2in]{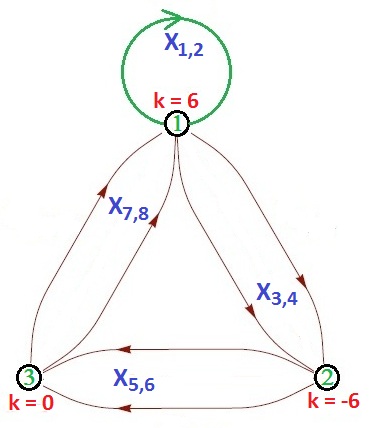}}
\caption{Quiver Diagram for Fano ${\cal B}_3$}
	\label{fig:b3fano}
\end{figure}
There are $N_T=4$ terms in $W$ for the $G=3$ node quiver with $E=8$ matter fields.
The quiver diagram can be constructed as shown in figure~\ref {fig:b3fano}.
Clearly, $N_T-E+G=-1$ and cannot admit tiling presentation. However,
the abelian $W$ is non-zero. So, it must be possible to do the 
forward algorithm after deducing the quiver diagram with CS levels 
from the inverse algorithm.

To obtain the CS levels on the three nodes of the quiver,
we need the matrix $P=K.T$:
\sid
P = \left(
\begin{array}{ccccccc}
 3 & 0 & 1 & 0 & 0 & 0 & 1 \\
 0 & 3 & 1 & 0 & 0 & 0 & 1 \\
 3 & 0 & 0 & 1 & 0 & 0 & 1 \\
 0 & 3 & 0 & 1 & 0 & 0 & 1 \\
 0 & 0 & 4 & 0 & 6 & 0 & 1 \\
 0 & 0 & 0 & 4 & 6 & 0 & 1 \\
 0 & 0 & 4 & 0 & 0 & 6 & 1 \\
 0 & 0 & 0 & 4 & 0 & 6 & 1 \\
\end{array}
\right)
\sidd
Using the $Q_D$ charge (\ref {db3}) and 
the $P$-matrix elements, we can obtain the projected charge 
$\Delta$ of the matter fields and find the possible quiver
as shown in figure~\ref {fig:b3fano} whose charge matrix $d$ is given below:
\begin{equation}
d = \left(
\begin{array}{c|cccc}
    & X_i(i=1,2) & X_i(i=3,4) & X_i(i=5,6) & X_i(i=7,8)   \\
\hline
a=1 &  0 &  1 & 0 & -1 \\
a=2 &  0 &  -1 &1 &  0  \\
a=3 &  0 & 0 & -1 &  1
\end{array}
\right)~,
\end{equation}
The CS levels of the 3-nodes are 
\begin{equation}
 k_1=6,  k_2=-6, k_3=0~.
\end{equation}
\subsubsection{Hilbert series evaluation for ${\cal B}_3$ Theory}
The total charge matrix for the $B_3$ theory is given by:
\begin{equation}
Q = \left(
\begin{array}{ccccccc}
 1 & 1 & 3 & 3 & -1 & -1 & -6 \\
 1 & 1 & 1 & 1 & 0 & 0 & -4 \\
 0 & 0 & 0 & 0 & 1 & -1 & 0
\end{array}
\right) \label {tcb3}
\end{equation}
The symmetry group for this theory is $SU(2)^2 \times {U(1)}^2$. Here, we have 7 
$p_{\alpha}$ fields. Let us denote the
R-charge fugacity associated with the $p_1$, $p_2$ be $s_1$, with $p_3$, $p_4$ be $s_2$ and with $p_5$, ${\tilde p}_5$ to be $s_3$ and 
$s_4$ respectively. Since the R-charge of the internal point $p_6$ is 0, the
corresponding fugacity is set to unity. Using the $Q$ matrix, the Hilbert series of 
the mesonic moduli space can be
written as:
\begin{eqnarray}
g^{mes}(s_1,s_2,s_3,s_4;{\cal B}_3)&= & 
\oint_{|z_1|=1} \left(\frac{\text{$dz_1$}}{2 \pi  \text{$\mathit{i}$$z_1$} }\right) 
 \oint_{|b|=1} \left(\frac{\text{$db$}}{2 \pi  \text{$\mathit{i}$$b$} }\right)
\oint_{|z_2|=1} \left(\frac{\text{$dz_2$}}{2 \pi  \text{$\mathit{i}$$z_2$} }\right)
\nonumber\\
~&~&
 \left\{\frac{1}{{\left(1-s_1z_1z_2\right)^2}\left(1-s_2z_1^3z_2\right)^2
\left(1-s_3\frac{b}{z_1}\right)}
\right.\nonumber\\
&~&\left.\frac{1}{
\left(1-s_4\frac{1}{bz_1}\right)\left(1-\frac{1}{z_1^6z_2^4}\right)
}\right\}
\end{eqnarray}
\begin{table}[t]
\begin{center}
  \begin{tabular}{ | c  | c | c | c | c | c | c |}
    \hline
          & $SU(2)_1$ & $SU(2)_2$ & $U(1)_R$ & $U(1)_B$ & $U(1)_q$  & fugacity        \\  \hline
    $p_1$ &   1       &     0     & 0.164    & 0        &   1       &  $s_1x_1q$      \\  \hline
    $p_2$ &  -1       &     0     & 0.164    & 0        &   1       &  $s_1q/x_1$     \\  \hline
    $p_3$ &   0       &     1     & 0.782    & 0        &  -3       &  $s_2x_2/q^3$   \\  \hline
    $p_4$ &   0       &    -1     & 0.782    & 0        &  -3       &  $s_2/(x_2q^3)$ \\  \hline  
    $p_5$ &   0       &     0     & 0.039    & 1        &   0       &  $s_3b$         \\  \hline
    ${\tilde p}_5$ &   0       &     0     & 0        &-1        &   4       &  $s_4q^4/b$    \\  \hline
    $p_6$ &   0       &     0     & 0        & 0        &   0       &  $1$            \\  \hline
  \end{tabular}
\end{center}
\caption{\textbf{Various charges of perfect matchings under global symmetry of 
${\cal B}_3$ theory}. In this table, $s_i$ is the fugacity of 
R-charges, $x_1$ and $x_2$ are weights of two $SU(2)$ symmetries, $b$ and $q$ are 
fugacities of $U(1)_B$ and $U(1)_q$ symmetries respectively.}
\label{tb3}
\end{table}
The Hilbert series for mesonic moduli space with the change
of variables as
$t_1=\left(s_1s_2^{1/3}\right)$ and $t_2= \left(s_2^{4/3}s_3s_4\right)$ turns out to be
\begin{eqnarray}
g^{mes}(t_1,t_2;{\cal B}_3)&=&\frac{1}{\left(1-t_2^3\right)^2\left(1-t_1^3\right)^3} 
                         \left\{1+5 t_1^3+9 t_1^2 t_2-3 t_1^5 t_2+8 t_1 t_2^2+t_1^4 t_2^2-3 t_1^7 t_2^2
\right.\nonumber\\
~&~&~~\left.
+3 t_2^3-t_1^3 t_2^3-8 t_1^6 t_2^3+3 t_1^2 t_2^4-9 t_1^5 t_2^4-5 t_1^4 t_2^5-t_1^7 t_2^5\right\}~.
\end{eqnarray}
which is the form expected for toric CY 4-folds with two $U(1)$ symmetries.
Suppose, $R_1$ and $R_2$ be the R-charges corresponding to the fugacities $t_1$ and $t_2$. 
That is, $t_1 = e^{-\mu R_1}$ and $t_2 = e^{-\mu R_2}$, where 
$\mu$ is the chemical potential for the R-charge. 
Similar to the volume minimisation done for earlier cases, we  can find the value 
$\hat R_1, \hat R_2$ which minimises the volume $V({\cal B}_3)$. 
For the given $W$ (\ref {potb3}), we find that $R_1+R_2=1/3$.
Substituting this relation in the volume of ${\cal B}_3$ and minimising it, 
we get $R_1 =\frac{1}{24} \left(-3+\sqrt{57}\right)=0.189 $ and $R_2 = \frac{1}{24} \left(11-\sqrt{57}\right)=0.144$. The
R-charges of each $p_{\alpha}$  can be 
determined which we tabulate in Table~\ref{tb3}.

\section{Summary and open problems}
In this paper, we have systematized inverse algorithm by 
understanding the pattern of the charge assignments $Q_F,Q_D$ 
obtained for 14 Fano 3-folds from forward/tiling algorithm \cite{han8}.
Particularly, we used the second Betti number, symmetry of the 
CY 4-folds to fix the number of rows of $Q_D$ charge matrix
and the entries of both $Q_F$ and $Q_D$ matrix. Our ansatz
in section 4.1,   
which states that the rank of the non-abelian symmetry of $Q_F$ is 
one higher than that of $Q_D$, indicates the multiplicity of
which points in the toric data could be taken. Using the
ansatz, we took a possible choice of $Q_F,Q_D$ 
and performed the sequence of steps to obtain the 
quiver diagram, superpotential and the Chern-Simons levels.

The quivers for Fano ${\BP}^3,{\cal B}_1,{\cal B}_2$ and ${\cal B}_3$
as shown in figures \ref{fig:afano}--\ref{fig:b3fano} with the 
appropriate $W$ constructed from $K$-matrix showed that they 
cannot admit tiling presentation. It appears that forward algorithm
from the quiver data for ${\cal B}_2$ and ${\cal B}_3$ can be 
done to confirm the toric data.

Our charge assignments $Q_F,Q_D$ for the four Fano 3-folds
gives the expected form for the Hilbert series of the mesonic moduli space. 
So, we believe that our ansatz must be correct. 
Using the volume minimisation, we have tabulated the $R$-charge of
the fields $p_{\alpha}$'s. We obtained
the correct genus $g=33$ for ${\BP}^3$ Fano. However, it is not clear
why the genus computation for ${\cal B}_2$ is giving $g=33$ instead of 
$g=29$ \cite{han8}.

It will be interesting to do the higgsing approach \cite {tapo}
for CS quivers which does not admit tiling presentation. We hope 
to report on the higgsing procedure in a future publication.
In ref.~\cite{jpark}, the 
tiling rules for $SO/Sp$ quivers corresponding to 
the orientifolds of the CY 3-folds were proposed.
It is a challenging problem  to generalise the tiling for
orientifolds of the CY 4-folds
\cite{forc}
corresponding to
$SO/Sp$ CS quivers.

\vskip.5cm
\noindent
{\bf Acknowledgments}: We would like to thank Tapobrata Sarkar and 
Prabwal Phukon for discussions during the initial stages of this project.
P.R would like to thank the hospitality of Center for Quantum
spacetime, Sogang University where this work was done during
the sabbatical visit. This work was supported by the 
National Research Foundation of Korea(NRF) grant funded by the
Korea government(MEST) through the Center for Quantum Spacetime(CQUeST)
of Sogang University with grant number 2005-0049409. 
 

\vskip.5cm
\newpage

\begin{thebibliography}{99}
\bibitem{bag}J.Bagger and N.Lambert, ``Modeling multiple M2's,'' 
Phys. Rev. {\bf D75}, 045020 (2007) [arXiv:hep-th/0611108]. ``Gauge symmetry and
Supersymmetry of Multiple M2-Branes,'' Phys. Rev. {\bf D77},065008 
(2008) [arXiv:0711.0955[hep-th]]. ``Comments on Multiple M2-branes,'' JHEP{\bf 0802}, 105
(2008) [arXiv:0712.3738[hep-th]].
\bibitem{gus}A.Gustavsson, ``Algebraic structures on parallel M2-branes,''  Nucl.Phys. {\bf B811}, 66 (2009)
[arXiv:0709.1260[hep-th]]. ``Selfdual strings and loop space Nahm equations,'' JHEP {\bf 0804}, 083 (2008)
[arXiv:0802.3456[hep-th]].
\bibitem{ram} M.Van Raamsdonk, ``Comments on the Bagger-Lambert theory and multiple
M2-branes,'' JHEP {\bf 0805}, 105 (2008) [arXiv:0803.3803].
\bibitem{abjm}O.Aharony, O. Bergman, D.L. Jafferis and J. Maldacena,
``N=6 superconformal Chern-Simons-matter theories, M2-branes and their gravity duals, ''JHEP {\bf 0810},091 (2008)
[arXiv:0806.1218[hep-th]].
\bibitem{kleb}I.R.Klebanov and G.Torri, ``M2-branes and AdS/CFT,'' Int.J.Mod.Phys. {\bf A25} 332 (2010) [arXiv:0909.1580[hep-th]].
\bibitem{mar}D.Martelli and J. Sparks, ``Moduli spaces of Chern-Simons
quiver gauge theories and AdS$_4$/CFT$_3$,'' Phys.Rev. {\bf D78}, 126005 (2008) [arXiv:0808.0912[hep-th]].
\bibitem{han1}B.Feng, A.Hanany and Y.H.He, ``D-brane gauge theories from toric singularities 
and toric duality,'' Nucl. Phys. {\bf B595}, 165 (2001) [arXiv:hep-th/0003085].
\bibitem{ken}A. Hanany and K.D. Kennaway, ``Dimer models and toric diagrams,''
[arXiv:hep-th/0503149].
\bibitem{he1}S.Franco, A.Hanany, K.D. Kennaway,D.Vegh and B.Wecht,
``Brane dimers and quiver gauge theories,'' JHEP {\bf 0601}, 096 (2006) [arXiv:hep-th/
0504110].
\bibitem{han3}A.Hanany and A.Zaffaroni, ``Tilings, Chern-Simons Theories and M2 branes,'' JHEP {\bf 0810}, 111 (2008)
[arXiv:0808.1244[hep-th]].
\bibitem{yama}K.Ueda and M.Yamazaki, ``Toric Calabi-Yau four-folds dual to Chern-Simons-matter
theories,'' JHEP {\bf 0812}, 045 (2008) [arXiv:0808.3768[hep-th]].
\bibitem{han4}A.Hanany, D.Vegh, A.Zaffaroni, ``Brane Tilings and M2 branes,'' JHEP {\bf 0903}, 012 (2009) [arXiv:0809.1440].
\bibitem{han5}S.Franco, A.Hanany, J.Park and D.Rodriguez-Gomez,``Towards M2-brane 
Theories for Generic Toric Singularities,''JHEP {\bf 0812}, 110 (2008) [arXiv:0809.3237
[hep-th]].
\bibitem{han6}A.Hanany and Y.H.He, ``M2-branes and Quiver Chern-Simons: A Taxonomic Study,''
arXiv:0811.4044[hep-th].
\bibitem{han7a}J.Davey, A.Hanany, N.Mekareeya and G.Torri, ``Phases of M2-brane Theories,'' JHEP {\bf 0906}, 025 (2009)
[arXiv:0903.3234[hep-th]].
\bibitem{han7b} J.Davey, A.Hanany, N.Mekareeya and G.Torri, ``Higgsing M2-brane Theories,''
arXiv:0908.4033[hep-th].
\bibitem{han8}J.Davey, A.Hanany, N.Mekareeya and G.Torri, ``M2-Branes and Fano 3-folds,''
arXiv:1103.0553[hep-th].
\bibitem{tapo}P. Agarwal, P. Ramadevi, T. Sarkar, ``A note on dimer models and D-brane gauge 
theories,'' JHEP {\bf 0806} 054 (2008) [arXiv:0804.1902].
\bibitem{sang}Sangmin Lee,``Superconformal field theories from crystal lattices,''
Phys.Rev.{\bf D75} 101901 (2007)[arXiv:hep-th/0610204].
\bibitem{sang1}Sangmin Lee, Sungjay Lee and Jaemo Park, ``Toric AdS4/CFT3 duals and M-theory Crystals,''
JHEP {\bf 0705} 004 (2007) [arXiv:hep-th/0702120].
\bibitem{jpark}Sebastian Franco, Amihay Hanany, Daniel Krefl, Jaemo Park, Angel M. Uranga and David Vegh, `` Dimers and Orientifolds'', JHEP
{\bf 0709}, 075 (2007) [arXiv:0707.0298[hep-th]].
\bibitem{forc}Davide Forcella, Alberto Zaffaroni, `` N=1 Chern-Simons theories, orientifolds and Spin(7) cones'',
 JHEP {\bf 1005}, 045 (2010) [arXiv:0911.2595[hep-th]].
\end{thebibliography}
\end{document}